\documentclass[prd,11pt,reprint,onecolumn,superscriptaddress,amsfonts,nofootinbib]{revtex4-2}

\usepackage{amssymb,amsmath,amsthm}
\usepackage{graphicx}
\usepackage{slashed}
\usepackage{ulem}

\newcommand{\Tr}{\mathrm{Tr}\, }
\newcommand{\ii}{\mathrm{i}}
\newcommand{\dd}{\mathrm{d}}

\begin{document}
\title{Non-topological fractional fermion number in the Jackiw-Rossi model}
\author{Caio Almeida}
\affiliation{Graduate Program in Physics (PPG-FIS), Universidade Federal do ABC, Santo Andr\'e, S.P., Brazil}
\author{Alberto Alonso-Izquierdo}
\affiliation{Departamento de Matematica Aplicada, University of Salamanca, Spain}
\affiliation{Instituto de Fisica Fundamental y Matematicas, IUFFyM, University of Salamanca, Spain}
\author{Rodrigo Fresneda}
\affiliation{CMCC-Universidade Federal do ABC, Santo Andr\'e, S.P., Brazil}
\author{Juan Mateos Guilarte}
\affiliation{Departamento de Fisica Fundamental, University of Salamanca, Spain}
\affiliation{Instituto de Fisica Fundamental y Matematicas, IUFFyM, University of Salamanca, Spain}
\author{Dmitri Vassilevich}
\email{dvassil@gmail.com}
\affiliation{CMCC-Universidade Federal do ABC, Santo Andr\'e, S.P., Brazil}
\affiliation{Physics Department, Tomsk State University, Tomsk, Russia}

\begin{abstract}
We compute the vacuum fermion current in $(2+1)$ dimensional Jackiw-Rossi model by using the $1/m$ expansion. The current is expressed through a weighted $\eta$-function with a matrix weight. In the presence of such a weight, the usual proof of topological nature of $\eta(0)$ is not longer applicable. Direct computations confirm the following surprising result: the fermion number induced by vortices in the  Jackiw-Rossi model is \textit{not} topological.
\end{abstract}

\maketitle

\section{Introduction}\label{sec:Intro}
As we know from the pioneering paper by Jackiw and Rebbi \cite{Jackiw:1975fn} and from the subsequent development reported in \cite{Niemi:1984vz}, the fermion number of solitons can take fractional and even irrational values. In the known cases, the fermion number is topological. This means that it depends on the boundary or asymptotic values of the background fields and is not sensitive to smooth variations of these fields in the interior of manifold. From the very beginning, the fermion number fractionization had applications to condensed matter physics \cite{Jackiw:1981wc}. More recently, this mechanism was applied to the physics of topological insulators, see \cite{Qi:2011zya}.  

Among the planar ($2+1$-dimensional) solitonic systems, a prominent role is played by the Abrikosov-Nielsen-Olesen (ANO) vortex. There are many possible ways to couple fermions to this system, and thus there are many quantum systems which include the ANO vortex as a bosonic sector. In a supersymmetric model, the one-loop shift of the mass of the vortex was calculated in \cite{Vassilevich:2003xk,Rebhan:2003bu}. In a pure bosonic model, this was done in \cite{AlonsoIzquierdo:2004ru,izquerdo2005,izquierdo2016} while (non-supersymmetric) fermions were added in \cite{bordag2003,Graham:2004jb}. 

The fermion number fractionization in $2+1$ dimensions on a pure gauge field background was calculated in \cite{Niemi:1983rq}. For a singular magnetic vortex this effect was considered in \cite{Sitenko:1996np}. For a pair of fermions coupled to both gauge and Higgs fields of the ANO vertex, the half-integer fermion fractionization was obtained in \cite{Chamon:2007pf,Chamon:2007hx} (see also the preceding papers \cite{Hou:2006qc,Jackiw:2007rr}). In these models, the fermions have the elementary electric charge $e$ while the scalar fields possess charge $2e$. Thus, the ANO vortex gets a fractional flux. There is a way of coupling a single generation of fermions to the ANO system which is given by the Jackiw-Rossi model \cite{Jackiw:1981ee}. This coupling reminds us of planar superconducting systems. A candidate for the fractional flux vortex in such systems was recently found experimentally \cite{Tanaka:2018}. This discovery motivated a study \cite{Yanagisawa:2020} of fermion charge fractionization in the Jackiw-Rossi model. The computations in this paper were based on the usual relation between vacuum fermion number and the $\eta$-function of the Hamiltonian which is not correct in the Jackiw-Rossi model, as will be demonstrated below.

The purpose of this paper is to analyze the vacuum fermion number in the Jackiw-Rossi model paying special attention to its topological (or rather non-topological) nature.  First, we observe that the interaction between scalar and spinor fields does not allow to immediately relate the fermion number to the $\eta$-function of an operator. One has to double the spinor components. This is similar to what has been done in \cite{Jackiw:1981ee,Weinberg:1981eu} to analyses the zero-energy spectrum, but we do this in the path integral formalism following the method of \cite{Novozhilov:1994xs}. The fermion density is then related to an $\eta$-function which, however, is weighted with a matrix. The presence of this matrix destroys the standard proof of vanishing local variations of $\eta$ and of the topological nature of this quantity. We go on by computing a few leading term in the large mass expansion of the fermion number with the heat kernel methods and confirm the presence of non-topological contributions depending on the profiles of magnetic and Higgs fields rather than on their global characteristics. 

This paper is organized as follows. In the next section, we derive an expression for the vacuum fermion number in Jackiw-Rossi model in terms of an $\eta$ function with a matrix weight. Section \ref{sec:Heat} is dedicated to the heat kernel evaluation of fermion number. First, we show why the standard proof of topological nature does not work for weighted $\eta$ function. Then, we pinpoint the non-topological fermion number within the large mass expansion. Some concluding remarks are given in the last section.

\section{Fermion number in the Jackiw-Rossi model}
The Jackiw-Rossi model \cite{Jackiw:1981ee} in $(2+1)$ dimensions is described by the Lagrangian
\begin{equation}
\mathcal{L}=\bar \psi (\gamma^\mu (\ii \partial_\mu -eA_\mu))\psi -\frac 12 \ii g \phi \bar\psi\psi^C +\frac 12 \ii g^*\phi^* \bar\psi^C \psi -m\bar\psi\psi \label{JRL}
\end{equation}
governing the dynamics of a two complex component spinor field $\psi$ coupled to a gauge and a complex scalar fields $A_\mu$, $\phi$.

As compared to the original work \cite{Jackiw:1981ee} a mass term has been added.
For convenience, let us take the $\gamma$-matrices in Majorana representation:
\begin{equation}
\gamma^0=\sigma^2,\qquad \gamma^1=\ii \sigma^1,\qquad \gamma^{2} = \ii \sigma^3.\label{JRgam}
\end{equation}
Then the charge conjugation matrix can be taken as $C=-\gamma^0$. We have the usual relations $C\gamma^\mu C^{-1}=-\gamma^{\mu\, T}$, $\psi^C=\psi^*$, etc.

We assume that bosonic fields $A_\mu$ and $\phi$ belong to the topological class of an ANO vortex. This configuration is static, so that $A_0=0$ and all fields do not depend on time. We are not going to use the exact profile functions, though it will be important to us that this configuration is localized somewhere near the origin. If $r$ is the radial coordinate, for $r\to\infty$ we have
\begin{equation}
|\phi|\to v,\qquad D_j\phi\to 0,\qquad F_{jk}\to 0 .\label{asANO}
\end{equation}
Here and in what follows $x^j$, $x^k$, etc denote spatial coordinates. $D_j\phi =(\partial_j +2\ii e A_j)\phi$ is a gauge covariant derivative, depending on the charge of the field it acts upon; therefore, in our notation, $D_j \phi^{*} \equiv (D_j\phi)^{*}$. Note that the electric charge of $\phi$ is $2e$, $F_{jk}\equiv \partial_jA_k-\partial_kA_j$, and $v$ is a minimum of the Higgs potential. All functions in (\ref{asANO}) go to their asymptotic values exponentially fast. Let $\mathcal{N}\in\mathbb{Z}$ be the topological charge of the vortex. The magnetic flux quantization condition
\begin{equation}
\frac e{\pi}\int d^2x\, F_{12}=\mathcal{N} \label{flux}
\end{equation}
has an unusual factor on the right hand side due to the charge $2e$ of $\phi$. This is why we say that the vortex has a fractional flux.

The Lagrangian (\ref{JRL}) besides the $\bar\psi\psi$ contains also the  $\bar\psi \psi^*$ and $\bar\psi^*\psi$ couplings to the Higgs field. Thus, it does not have the form that allows to relate immediately the states to eigenfunctions of some differential operator. To overcome this difficulty we pass to doubled spinors following the approach developed in the paper \cite{Novozhilov:1994xs} (see also \cite{Ball:1989hn,Novozhilov:1994he}). We introduce
\begin{equation}
\Psi:=\begin{pmatrix} \psi \\ \bar\psi^T \end{pmatrix}. \label{JRPsi}
\end{equation}
 With the help of identities
\begin{eqnarray*}
&&\int d^3x \bar \psi \gamma^\mu (\ii \partial_\mu -eA_\mu)\psi =
\int d^3x \psi^T \gamma^{\mu\, T} (\ii \partial_\mu +eA_\mu)\bar\psi^T,\\
&&\bar\psi^C\psi=\psi^T\gamma^0\psi,\qquad \bar\psi\psi^C=-\bar\psi \gamma^0\bar\psi^T,\qquad -m\bar\psi\psi =m\psi^T \bar\psi^T
\end{eqnarray*}
we rewrite the action as
\begin{equation}
S=\frac 12 \int d^3x \Psi^T \widehat F \Psi \label{JRS}
\end{equation}
with
\begin{equation}
\widehat F=\begin{pmatrix}
\ii g^*\phi^* \gamma^0 & \gamma^{\mu\, T} (\ii\partial_\mu + eA_\mu)  +m \\
\gamma^\mu(\ii\partial_\mu -eA_\mu) -m & \ii g\phi \gamma^0
\end{pmatrix}.
\end{equation}
The corresponding Hamiltonian reads
\begin{equation}
H=\begin{pmatrix}
\alpha^j (\ii \partial_j -eA_j) - \beta m & \ii g \phi \\
-\ii g^* \phi^* & -\alpha^j (\ii \partial_j + eA_j) -\beta m
\end{pmatrix}. \label{JRH}
\end{equation}
Here, as usual, $\beta\equiv \gamma^0$ and $\alpha^j=\beta\gamma^j$.

Let us consider the effective action $W$ which is obtained by integrating out the fermionic degrees of freedom,
\begin{equation}
e^{\ii W}=\int\mathcal{D}\psi\, \mathcal{D}\bar\psi\, \exp\left( \ii \int d^3 x \mathcal{L}\right).\label{defW}
\end{equation}
This action depends on the background bosonic fields $\phi$ and $A_\mu$. The charge density is given by the variational derivative
\begin{equation}
j^0=-\frac 1e \, \frac{\delta W}{\delta A_0}.\label{j0}
\end{equation}
The same effective action $W$ can be written through a path integral over the doubled spinors $\Psi$ as
\begin{equation}
W=-\ii \, \ln \int\mathcal{D}\Psi\, \exp\left( \ii \tfrac 12 \int d^3x\, \Psi^T \widehat F \Psi \right)=-\frac {\ii}2 \ln\det (\widehat F),\label{WF}
\end{equation}
see \cite{Novozhilov:1994xs}. The functional integration measure became $\mathcal{D}\Psi =\mathcal{D}\psi\, \mathcal{D}\bar\psi$.

Symbolically, we may write
\begin{equation}
j^0=\frac{\ii}{2e} \, \mathrm{Tr}\left( \frac{\delta\widehat F}{\delta A_0} \, \widehat{F}^{-1} \right).\label{j01} 
\end{equation}
To give precise meaning to this formula one has to invert $\widehat F$ and regularize the functional trace. 

After having calculated the variational derivative in (\ref{j01}) one puts the background fields to their values for the static vortex configuration. On such a background, the eigenfunctions of Hamiltonian (\ref{JRH}) can be taken depending on the spatial coordinates $\vec{x}$ only,
\begin{equation}
H\Psi_n (\vec{x})=E_n\Psi_n (\vec{x}). \label{HPsin}
\end{equation}
The energy spectrum has both discrete and continuous parts. To avoid notation clutter we write the formulas below as if the whole spectrum were discrete. 

For $g=0$, the Hamiltonian $H$ consists of two hermitian anticommuting parts. Thus one can easily show that $(H(g=0))^2\geq m^2$. Consequently, if $|m|>|g\phi|$ the full Hamiltonian does not have zero energy eigenstates.

The vectors 
\begin{equation}
\Psi_{\omega,n}(\vec{x},t)=(2\pi)^{-1/2}e^{-\ii \omega t} \, \Psi_n(\vec{x})
\end{equation}
form a basis for the space of square integrable 4-spinors on $\mathbb{R}^3$. To compute  (\ref{j01}) one has to sandwich the expression under the trace between $\Psi_{\omega,n}^\dag$ and $\Psi_{\omega,n}$, integrate over $\omega$ and sum over $n$. To regularize the $\omega$-integral we use a symmetric time-splitting regularization. Namely, we take two eigenvectors at shifted time arguments, $\Psi_{\omega,n}^\dag(\vec{x},t)$ and $\Psi_{\omega,n}(\vec{x},t+\Delta t)$. After computing the integral, we take the limits $\tfrac 12(\lim_{\Delta t\to +0} + \lim_{\Delta t \to -0})$. To regularize the sum, we multiply the expression by $|E_n|^{-s}$ with $\Re s$ sufficiently large to ensure the convergence and analytically continue to $s= 0$ afterwards.
\begin{eqnarray}
&&j^0(\vec{x},t)=-\frac{\ii}2 \frac 12 \left( \lim_{\Delta t\to +0} + \lim_{\Delta t \to -0} \right) \int_{-\infty}^\infty \frac{d\omega}{2\pi}\, \sum_n |E_n|^{-s} \nonumber \\
&&\qquad\qquad \times \Psi_n^\dag (\vec{x}) \begin{pmatrix}
1 & 0 \\ 0 & -1 \end{pmatrix} \Psi_n(\vec{x}) \frac{e^{-\ii \omega\, \Delta t}}{\omega +\ii\, 0 \, \mathrm{sgn}(\omega) -E_n }\,.\label{j02}
\end{eqnarray}  
After performing the integration over $\omega$ one obtains
\begin{equation}
j^0(\vec{x},t)=-\frac 14 \sum_n \Psi_n^\dag (\vec{x}) \begin{pmatrix}
1 & 0 \\ 0 & -1 \end{pmatrix} \Psi_n(\vec{x})\, \mathrm{sgn}\, (E_n)\, |E_n|^{-s} \label{j03}
\end{equation}
The analytic continuation to $s=0$ is understood in both formulas (\ref{j02}) and (\ref{j03}).

Let $Q$ be a smooth bounded matrix-valued function (a smooth endomorphism). The $\eta$ function of $H$ smeared with $Q$ is defined as
\begin{equation}
\eta(s,H;Q)=\Tr \left(Q\, \mathrm{sgn}\, (H)\, |H|^{-s} \right)=
\Tr \left( Q \cdot (H^2)^{-s/2} \cdot (H/|H|) \right)=\Tr \left( Q \cdot (H^2)^{-\frac{s+1}2} H\right). \label{eta}
\end{equation}
Here again $s$ is a complex parameter. The trace in (\ref{eta}) exists if $\Re\, s$ is sufficiently large. This function can be analytically continued as a meromorphic function to the whole complex plane. At $s=0$, equation (\ref{j03}) yields
\begin{equation}
\int d^2x\, j^0(x)\, \rho(x)=-\frac 14\, \eta(0,H;\rho \tau_3)\,, \label{j0eta}
\end{equation}
where $\rho$ is a smooth localizing function of compact support, and  $\tau_3=\left(\begin{array}{cc}1 & 0\\ 0 &-1\end{array}\right)$. An integrated version of (\ref{j03}) gives an expression for the fermion number $N$ in terms of the $\eta$ function, 
\begin{equation}
N\equiv \int d^2x\, j^0(x)=-\frac 14\, \eta(0,H; \tau_3). \label{Neta}
\end{equation}
There are two important differences from the corresponding formula derived in the seminal paper \cite{Niemi:1983rq}. These are the coefficient $1/4$ instead of $1/2$ and the presence of $\tau_3$ in the $\eta$ function. Both are caused by our spinor field doubling procedure. The presence of $\tau_3$ has a profound consequence: the standard proof that $\eta(0)$ is topological in 2D does not work any more.

\section{Heat kernel computations of the fermion current}\label{sec:Heat}
\subsection{Why the standard proof of $N$ being topological does not work for the JR model}\label{sec:why}
Here we study local variations of the $\eta$-function with and without a matrix weighting factor. Our method goes back to the paper by Atiyah, Patodi and Singer\cite{Atiyah:1980jh}. We closely follow the procedure presented in \cite{Gilkey:book}. A slightly different method was used in \cite{AlvarezGaume:1984nf}. 

Let $H(\varepsilon)=H+\varepsilon\, h$ where $h$ is a perturbation caused by an infinitesimal localized variation of background bosonic fields $\phi$ and $A$. Let us consider the case $Q=1$. By using Lemma 1.10.2 of \cite{Gilkey:book} we can express the variation of the $\eta$ function through a $\zeta$ function weighted with $h$
\begin{equation}
\left.\frac{\dd}{\dd\varepsilon}\right\vert_{\varepsilon=0} \eta(s,H(\varepsilon)) =
-s \Tr \left[  h\, (H^2)^{-\frac{s+1}2} \right]=-s\zeta\left(\frac{s+1}2 ,H^2;h\right).\label{deleta}
\end{equation}
Now, we need some basics on the spectral functions. Let $L$ be a Laplace type operator on a manifold $M$ of dimension $n$ with or without boundary. Let $h$ be a smooth matrix valued function. Then, residues of the $\zeta$ function can be expressed by the formula 
\begin{equation}
\mathrm{Res}_{u=\frac{n-k}2} \left( \Gamma(u)\zeta(u,L;h)\right)=a_k(L;h) \label{Res}
\end{equation}
through the heat kernel coefficients defined through the following asymptotic expansion at $t\to +0$
\begin{equation}
\Tr \left( h e^{-tL}\right)\simeq \sum_{k=0}^\infty t^{\frac{k-n}2} a_k(L;h). \label{hkex}
\end{equation}
By Eq.\ (\ref{deleta}), the derivative $(\dd \eta(0,H))/(\dd\varepsilon)\vert_{\varepsilon=0}$ is given by the residue of $\zeta(u,H^2;h)$ function at $u=\tfrac 12$ which is in turn proportional to $a_1(H^2;h)$. Since $h$ is localized inside the manifold and does not extend to boundaries or asymptotic regions, the coefficient $ a_1(H^2;h)$ vanishes. We conclude that $\eta(0,H)$ does not change under local variations $H\to H(\varepsilon)=H+\varepsilon h$ and thus is a topological invariant.

The key point of the proof presented above was the Eq.\ (\ref{deleta}) relating the variation of the $\eta$ function to a residue of the $\zeta$ function which happened to be local and vanishing in the dimension $n=2$. Roughly speaking, to get (\ref{deleta}) one needs to differentiate $\eta(0,H)$ as if it were a usual function of a commutative variable. This property is ensured by the possibility of reordering operators under the trace. This possibility is (partially) lost if $Q$ does not commute with $H$. In such a case, the variation of $\eta(0,H;Q)$ cannot be written in the simple form of (\ref{deleta}) and all subsequent arguments break down.

\subsection{Computations for the JR model}
To evaluate the large mass expansion of the current (\ref{j0eta}) we shall use the method proposed in \cite{Alonso-Izquierdo:2019tms,MateosGuilarte:2019eem}. 
With the help of the identity
\begin{equation}
\int_0^\infty dt\, t^a e^{-bt}=b^{-(1+a)}\Gamma(1+a) \label{iden} 
\end{equation}
we write
\begin{equation}
\eta(s,H,\rho\tau_3)=\frac 1{\Gamma\left( \frac{s+1}2 \right)} \int_0^\infty dt\, t^{\frac{s-1}2}\Tr \left(\rho\tau_3He^{-tH^2} \right).\label{etasH}
\end{equation}
Let us introduce a shifted operator $H_\rho=H-\varepsilon\rho\tau_3$ with $\varepsilon$ being a real parameter. Then
\begin{equation}
\eta(s,H,\rho\tau_3)=\frac 1{2\Gamma\left( \frac{s+1}2 \right)} \int_0^\infty dt\, t^{\frac{s-3}2} \left.\frac{\dd}{\dd\varepsilon}\right\vert_{\varepsilon=0} \Tr \left(e^{-tH_\rho^2} \right).\label{eta21}
\end{equation}
To evaluate this expression by using a large mass expansion we isolate $m^2$ in $H_\rho^2$ and take the limit $s\to 0$ to obtain
\begin{equation}
\eta(0,H,\rho\tau_3)=\frac 1{2\sqrt{\pi}} \int_0^\infty dt\, t^{-\frac{3}2} \left.\frac{\dd}{\dd\varepsilon}\right\vert_{\varepsilon=0}e^{-tm^2} \Tr \left(e^{-t\tilde H_\rho^2} \right), \label{eta22}
\end{equation}
where $\tilde{H}_\rho^2\equiv H_\rho^2 -m^2$. Next, we make the heat kernel expansion (\ref{hkex}) and integrate over $t$. 
\begin{eqnarray}
\eta(0,H;\rho\tau_3)&\simeq& \frac 1{2\sqrt{\pi}} \int_0^\infty dt\,\sum_{k=0}^\infty t^{\frac{k-5}2}\left. \frac{\dd}{\dd\varepsilon}\right\vert_{\varepsilon=0}a_k(\tilde H_\rho^2)\, e^{-tm^2}\nonumber\\
&=&\frac 1{2\sqrt{\pi}}\sum_k \Gamma\left( \frac{k-3}2\right)\, |m|^{3-k} \left.\frac{\dd}{\dd\varepsilon}\right\vert_{\varepsilon=0}a_k(\tilde{H}_\rho^2).\label{JRetaex}
\end{eqnarray}
Here $a_k(\tilde H_\rho^2)\equiv a_k(\tilde H_\rho^2;1)$. The integral above is convergent if the contributions of heat kernel coefficients $a_k$ with $k\leq 3$ vanish. We shall check this condition below.

To be able to use universal expressions for the heat kernel coefficients (see, e.g.,  \cite{Vassilevich:2003xt}) we represent the operator $\tilde{H}_\rho^2$ in the canonical form 
\begin{equation}
\tilde{H}_\rho^2=-(\nabla_j\nabla_j +E), \label{canH}
\end{equation}
where $\nabla_j=\partial_j+\omega_j$ plays the role of a covariant derivative while $E$ is a matrix valued potential. For our operator they read
\begin{eqnarray}
&&E=\begin{pmatrix}
\frac e2 \beta\, \epsilon^{jk}F_{jk}-|g\phi|^2 & g\alpha^jD_j\phi +2\ii \beta g \phi m\\
g^*\alpha^jD_j\phi^* -2\ii \beta g^* \phi^*m & -\frac e2 \beta\, \epsilon^{jk}F_{jk}-|g\phi|^2
\end{pmatrix} -2\varepsilon\rho\beta m \begin{pmatrix}
1 & 0 \\ 0 & -1 
\end{pmatrix} \label{E} \\
&& \omega_j=\begin{pmatrix}
\ii e A_j & 0 \\ 0 & -\ii e A_j  
\end{pmatrix} +\ii \alpha^j \varepsilon\rho \label{omega}
\end{eqnarray}
In this section, we are working in a Euclidean space with a positive unit metric. We still keep the distinction between upper and lower indices of some quantities which have a $(2+1)$-dimensional origin. For example, $A$ always appears with a subscript, while $\alpha$ and $\gamma$ come with superscripts. The summation over repeated indices is always done with the Kronecker symbol independently of the position of indices. This prescription destroys the balance between upper and lower indices within formulas, but keeps the notations simple and unambiguous.

Each heat kernel coefficient $a_k$ is an integral of a trace of a local polynomial constructed from $E$, the field strength $\Omega_{ij}=[\nabla_i,\nabla_j]$, and their repeated covariant derivatives (e.g., $E_{;j}=[\nabla_j,E]$, etc). For example,
\begin{eqnarray}
&&E_{;j}=\begin{pmatrix}
\frac e2 \beta \partial_j(\epsilon^{kl}F_{kl}) -2\varepsilon \beta m\partial_j\rho -\partial_j |g\phi|^2 &
g\alpha^k (D_jD_k\phi)+2\ii g \beta m (D_j\phi) \\
g^*\alpha^k (D_jD_k\phi^*)-2\ii g^* \beta m (D_j\phi^*) &
-\frac e2 \beta \partial_j(\epsilon^{kl}F_{kl}) +2\varepsilon \beta m \partial_j\rho -\partial_j|g\phi|^2
\end{pmatrix}\nonumber\\
&&\qquad + \begin{pmatrix}
-\ii\gamma^j e \hspace{1.5pt} \epsilon^{kl}F_{kl} & 2g\beta \epsilon^{jk}(D_k\phi) +4\gamma^jmg\phi \\
2g^*\beta \epsilon^{jk}(D_k\phi^*) -4\gamma^jmg^*\phi^* & \ii\gamma^j e \hspace{1.5pt} \epsilon^{kl}F_{kl}  
\end{pmatrix}\varepsilon\rho \label{Ej}
\end{eqnarray}
and
\begin{equation}
\Omega_{ij}=\begin{pmatrix}
\ii e F_{ij} +\ii \varepsilon \left(\alpha^j (\partial_i\rho) - \alpha^i (\partial_j\rho)\right) & 0 \\ 0 & -\ii e F_{ij} +\ii \varepsilon  \left(\alpha^j (\partial_i\rho) - \alpha^i (\partial_j\rho)\right) \label{Oij}
\end{pmatrix}\,.
\end{equation}
All invariants entering $a_k$ have the canonical mass dimension $k$. On manifolds without boundaries, all coefficients with odd values of $k$ vanish.

By a direct computation with the expressions from \cite{Vassilevich:2003xt}, one obtains that the contributions of $a_0$ and $a_2$ to (\ref{JRetaex}) vanish thus fulfilling the consistency condition presented below (\ref{JRetaex}).

Let us compute the current as an expansion in $\phi$ and its derivatives keeping the terms up to $D^2$ and $\phi^2$. Since $[D_j,D_k]\propto F_{jk}$ we shall also keep the terms with $F$ and $F\phi^2$, while $m$ can enter with any power. 

It is important to establish upper bounds on the number $k$ of the heat kernel coefficient which contains the required invariants. Consider the term $\rho\epsilon^{jk}F_{jk}$. It has canonical mass dimension $3$. Thus, in $a_k$ it has to be multiplied by $m^{k-3}$. Since $\omega$ does not contain $m$, this requires product of $E$ or of its derivatives at least $k-3$ times -- an expression which has the mass dimension greater than or equal to $2(k-3)$. Since the mass dimension of $a_k$ is $k$, we have the upper bound $k\leq 6$. In a similar way one comes to the conclusion that $\rho (D\phi)(D\phi^*)$ and $\rho |g\phi|^2 F$ terms may appear for $k\leq 10$. By refining these arguments one can exclude a lot of possible terms in the expansion and even improve the bounds mentioned above. At any rate, with explicit expressions from \cite{Fliegner:1997rk} for flat space heat kernel coefficients up to $a_{12}$, the rest may be done by a Wolfram Mathematica script.

We obtain that, up to the order considered, just a few terms in the heat kernel expansion contribute. The result reads
\begin{eqnarray}
&&a_4=\frac 1{4\pi}\int d^2x\, \mathrm{tr}\left( \tfrac 12 E^2 +\dots \right)=-\frac 1{\pi} \int d^2x \, e\epsilon^{jk}F_{jk}\varepsilon\rho m +\dots ,\label{a4E2}\\
&&a_6=\frac 1{4\pi}\int d^2x\, \mathrm{tr}\left( \tfrac 16 E^3 -\tfrac 1{12}E_{;j}E_{;j}+\dots \right)=
\frac 1{\pi}\int d^2 x\, \varepsilon\rho m\, |g|^2 \epsilon^{jk}\left( \frac{\ii}3 (D_j\phi)(D_k\phi^*) +\frac{5e}3 |\phi|^2F_{jk} \right)+\dots , \label{a6E3}\\
&&a_8=\frac 1{4\pi}\int d^2x\, \mathrm{tr}\left( \tfrac 1{24} E^4 +\dots \right)=
-\frac 2{3\pi} \int d^2x\, em^3 |g\phi|^ 2 \varepsilon\rho\, \epsilon^{jk}F_{jk}+\dots \,,\label{a8E4}
\end{eqnarray}
where dots denote irrelevant terms.

Thus, in the approximation adopted here,
\begin{equation}
j^0=\frac 1{8\pi}\, \frac m{|m|} \left[ e\epsilon^{jk}F_{jk} -\frac {\ii|g|^2}{6m^2}\, \epsilon^{jk} (D_j\phi)(D_k\phi^*) - \frac{|g\phi|^2}{3m^2} e\epsilon^{jk}F_{jk}  \right].\label{j05}
\end{equation}
The integral of $j^0$ gives the vacuum fermion number
\begin{equation}
N=\frac{\mathcal{N}}4 \, \frac m{|m|} -\frac {|g|^2e}{48\pi\, m\,|m|}\int d^2 x\, |\phi|^2 \epsilon^{jk}F_{jk} .\label{Nfin}
\end{equation}
To obtain this expression we integrated by parts and used the asymptotic conditions (\ref{asANO}) together with the relation (\ref{flux}). The first term on right hand side of (\ref{Nfin}) describes the (expected) quarter-integer quantization of the fermion number in the absence of scalar field $\phi$. The second term depends on the profiles of $|\phi|^2$ and $F_{jk}$ in the interior of manifold and thus is not topological.

\section{Conclusions}\label{sec:con}
In this paper, we have expressed the vacuum fermion number of the Jackiw-Rossi model through an $\eta$-invariant of a matrix-weighted Hamiltonian. We have pinpointed the reason why the standard proof of the topological nature of the (fractional) fermion number fails and we have also explicitly computed  a non-topological contribution to this quantity.
   
We have computed the fermion current in just a few leading orders of the large mass expansion, since this was enough for our purposes. If needed, further terms can also be calculated with the help of the flat space heat kernel expansion from the paper \cite{Fliegner:1997rk}. One can also use resummations of the heat kernel, see \cite{Barvinsky:1990up,Avramidi:1997jy}.

The fact that the fermion number depends on the profiles of the magnetic field and of the Higgs field should have some consequences for condensed matter physics. We are not ready to go deeper into this subject. We just mention a potentially related work which studies, both theoretically and experimentally, the influence of non-uniformity of the magnetic field on Hall conductivity for various planar systems \cite{Schirmer_2020}.

Speaking about future prospects, we would also like to mention the work \cite{Bazeia:2020nmz} which studies relations between the parameters of solitons and the fermion spectrum. Probably, these results can be lifted to the quantum level to gain information about the fermion fractionization and other similar effects.

\begin{acknowledgments}
This work was done in the framework of an agreement between the S\~ao Paulo Research Foundation (FAPESP) and the University of Salamanca, project 2017/50294-1 (SPRINT). The work was supported in parts by the project 2016/03319-6 of FAPESP, by the grants 305594/2019-2 and 428951/2018-0 of CNPq. Besides, D.V. was supported by the Tomsk State University Competitiveness Improvement Program. AAI and JMG acknowledge the Junta de Castilla y Leon for partial financial support, under Grants No. BU229P18 and No. SA067G19. CA received a graduate scholarship directly provided by the Federal University of ABC and, hence, would like to thank the institution for the financial support.
\end{acknowledgments}

\end{document}